\documentclass[reprint,amsmath,amssymb,aps]{revtex4-1}
\usepackage{graphicx}
\usepackage{subcaption}
\usepackage{dcolumn}
\usepackage{bm}
\usepackage[
margin=1in,
]{geometry}
\usepackage{natbib}
\usepackage{float}
\usepackage{multirow}
\usepackage{array}

\newcommand{\La}{\Lambda}

\begin{document}

\preprint{APS/123-QED}

\title{Imprint of a Steep Equation of State in the growth of structure}

\author{Mariana Jaber-Bravo}
 \email{jaber@fisica.unam.mx}
\author{Erick Almaraz}
\email{ealmaraz@estudiantes.fisica.unam.mx}
\author{Axel de la Macorra}
%
\affiliation{%
Instituto de F\'isica, Universidad Nacional Aut\'onoma de M\'exico, \\A.P. 20-364 CDMX 01000, M\'exico.
}
\date{\today}

\begin{abstract}
	We study the cosmological properties of a dynamical of dark energy (DE) component determined by a Steep Equation of State (SEoS) $w(z)=w_0+w_i\frac{(z/z_T)^q}{1+(z/z_T)^q}$. The SEoS has a transition at $z_T$  between two pivotal values ($w_i, w_0$) which can be taken as an early time and present day values of $w$ and the steepness is given by $q$. We describe the impact of this dynamical  DE  at background and  perturbative level. The steepness of the transition has a better cosmological fit than a conventional CPL model with $w=w_0+w_a(1-a)$. Furthermore, we analyze the impact of
	steepness of the  transition in the growth of matter perturbations and structure formation. This is manifest in the linear matter power spectrum, $P(k)$, the logarithmic growth function, $f\sigma_8(z)$, and the differential mass function $dn/d\log M(z=0)$. The differences in these last  three quantities is at a percent-level using the  same cosmological baseline parameters in our SEoS and a  $\Lambda CDM$ model. However, we find  an increase in the power spectrum, producing a bump at $k\approx k_T$ with $k_T\equiv a_TH(a_T)$ the mode associated to the time of the steep transition ($a_T = 1/(1+z_T)$). %
   Different dynamics of DE  lead  to a different amount of DM at present time which has an impact in Power Spectrum and accordingly in structure formation. 
\end{abstract}

\maketitle


\section{\label{sec:intro}Introduction}

The  standard  $\Lambda$CDM paradigm is based on the assumptions of homogeneity and isotropy of the Universe at large scales, the validity of General Relativity and the cosmological constant term  as cause of the accelerated cosmic expansion.
Although it has been proven successful when tested against observations  it faces some major theoretical issues  such as the extreme fine-tuning problem known as the cosmological constant problem \citep{RevModPhys.61.1} which leads to the necessity of extending it.
Some candidates  include  scalar field models  or modifications of General Relativity   

Observational probes coming from different  physical phenomena such as the temperature and polarization of cosmic microwave background (CMB) \citep{planck2018}, the luminosity distance of supernovae \citep{Scolnic:2017caz} or the statistical signature of the baryonic acoustic oscillations (BAO) from galaxy surveys \cite{Beutler:2011hx,Ross:2014qpa,Padmanabhan2pc,Alam:2016hwk,Kazin:2014qga},  quasars \cite{Font-Ribera:2013wce,Delubac:2014aqe} or voids \cite{Liang:2015oqc}, have improved significantly over the years.

In this work we choose to focus on an effective model of a fluid with free parameters. In this, we  consider the dark energy (DE) contribution, $\rho_{DE}$ to be a perfect fluid so dissipative terms will not be present.  In this situation we describe the dynamics of this component through its equation of state, $w(z)$, defined by:
\begin{equation}
p_{DE} (\rho_{DE}) = w(z)\rho_{DE}(z)
\end{equation}
which can be parameterized to match observations. 

Several proposals for $w(z)$ can be found in the literature \citep{Chevallier:2000qy, Linder:2002et, Doran:2006kp, KraussJonesHuterer2007, Linder:2006ud, Rubin:2008wq, 2009ApJ:703:1374S, 2010PhRvD..81f3007M, Hannestad:2004cb, Jassal:2004ej, Ma:2011nc, Huterer:2000mj, Weller:2001gf, Huang:2010zra, Barboza:2008rh}. These proposals attempt to describe the dynamics of dark energy without assuming a particular theoretical model, but providing practical parametrizations that can be readily confronted against observations
In this approach, a cosmological constant solution can be modelled as a fluid with pressure \linebreak $p_{\Lambda} = -\rho_{\Lambda}$, which implies an equation of state $w=-1$. 
This landscape has recently been extended to cover the background expansion rate as prescribed by $f(R)$ theories \citep{Jaime:2018ftn}. 

The study of the perturbative regime could  potentially be used to discriminate between a cosmological constant and models with a negative pressure component from Modified gravity. 
In this pursue, ongoing and upcoming surveys such as eBOSS \citep{eboss}, DESI \citep{DESIref}, LSST \citep{2009arXiv0912.0201L} and EUCLID \citep{2011arXiv1110.3193L} will provide extremely precise measurements of the growth of structure in the Universe, which in turn, will allow to probe the nature of the cosmic acceleration mechanism.

Studying the effect of dynamical DE into the clustering at large scales is thus a relevant task for the cosmological community. 
In this work we present the implications that a steep transition in the DE EoS, $w(z)$ has in the growth of structure. 

This paper is organized as follows. In Section \ref{sec:methods} we describe our model for dark energy,  the cosmology chosen and the analytical treatment used for the perturbations.  Section \ref{sec:results} comprises our main results in the particular case of a smooth dark energy component and its impact on linear observables in the perturbative regime.
Our conclusions and outlook are covered in section \ref{sec:conclusions}. 
%

\section{\label{sec:methods}Methods}
\subsection{\label{subsec:seos}Steep Equation of State}

In a previous work \citep{Jaber:2017bpx} we presented a parametric form for $w(z)$ inspired in quintessence fields and tested its free parameters  with observations such as the Baryion Acoustic Oscillations  (BAO) peak measured in galaxies or in the Lymann-$\alpha$ forest, as well as the compressed  Cosmic Microwave Background likelihood \citep{mukherjee, planck15DE}, and the local determination of $H_0$ included in \citep{localhubble}.  

Our form for the equation of state is: 

\begin{equation}
	\label{eq:steepeos}
	w(z) = w_0 + (w_i - w_0) \frac{(z/z_T)^q}{1+(z/z_T)^q}
\end{equation}
which allows for a steep transition to take place at a pivotal redshift $z = z_T$ with a steepness modulated by the exponent $q$. For this reason we dubbed this equation ``SEoS" (from ``Steep Equation of State") in this work.

We notice that in the  case  where the transition is smooth and occurs at a particular redshift: $z_T = q = 1$, we recover a form for the equation of state known as the Chevallier-Polarski-Linder parametrization (CPL)  \citep{Chevallier:2000qy,Linder:2002et} which has been widely used in the literature,
\begin{eqnarray}
	w(z; z_T=1, q=1) & = &w_0 + (w_i - w_0) \frac{z}{1+z} \nonumber \\
										&= & w_0 + w_a (1-a),
\label{eq:cpl}
\end{eqnarray}
where $w_a\equiv w_i-w_0$ and we keep the convention $a_0 = 1$.

\begin{figure}[t]
	\centering
	\includegraphics[width=\linewidth]{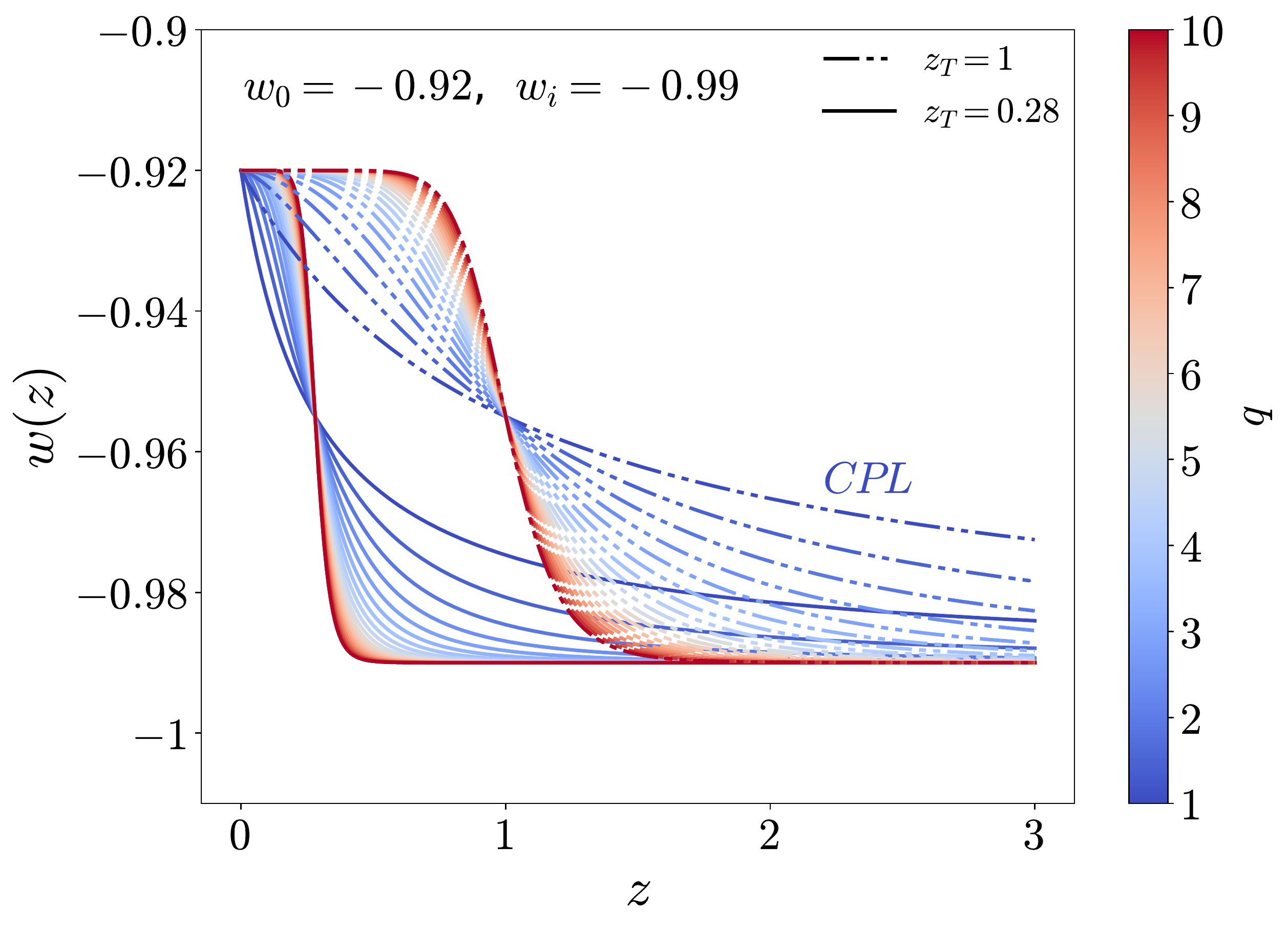}
	\caption{[Color online] Evolution of equation  \eqref{eq:steepeos} (``SEoS") for different values of the transition redshift, $z_T$, and the steepness parameter, $q$. The parameters $w_0$  and $w_i$ were fixed to $-0.92$ and $-0.99$, respectively, and $q$ was varied from $q=1$ (blue) to $q=10$ (red). Solid lines represent the evolution of $w(z)$ with $z_T=0.28$, and dot-dashed  lines indicate $z_T=1$. The CPL case  is explicitly labeled and corresponds to $z_T=q=1$.}
	\label{fig:eos} 
\end{figure}

In this work we will refer to the particular case of having arbitrary $w_0$ and $w_i$ but taking  $z_T = q = 1$, as the ``CPL limit'' of the SEoS \eqref{eq:steepeos}. 

We notice that the parameter $q$ modulates the steepness of the transition: the greater the value for $q$, the more abrupt is the transition, as figure \ref{fig:eos} shows. 

\subsection{\label{subsec:cosmology} Background models}
Once we have specified the equation for the dynamics of DE, the expansion rate (for a flat Universe) is given by:
\begin{equation}
	\label{eq:Hz}
	\frac{H(z)}{H_0}=\sqrt{\Omega_r^{(0)}(1+z)^4+\Omega_m^{(0)}(1+z)^3+\Omega_{DE}^{(0)}F(z)}
\end{equation}
where $H\equiv(\frac{da}{dt})(\frac{1}{a})$ is the Hubble parameter,   $t$  the  cosmic time, $a=(1+z)^{-1}$ the scale factor of the Universe and  $H_0 = 100\cdot h$ $km\cdot s^{-1} Mpc^{-1}$  the Hubble constant at present time. 
The   fractional densities of matter, radiation and dark energy at $z=0$, are given by $\Omega_m^{(0)}$, $\Omega_{r}^{(0)}$, $\Omega_{DE}^{(0)}$, respectively,  which follow the flatness constraint  $\Omega_m+\Omega_r+\Omega_{DE}=1$. 

The function $F(z)$ in equation \eqref{eq:Hz} encodes the evolution of the DE component in terms of its equation of state:
\begin{eqnarray}
	\label{eq:DE-f}
	F(z) & \equiv & \frac{\rho_{DE}(z)}{\rho_{DE}(0)} \\ \nonumber
	F(z) &=& \exp \left( -3 \int_0^z dz'\frac{1+w(z')}{1+z'} \right)
\end{eqnarray}


\begin{table*}
	\begin{center}
		\begin{tabular}{l |c c c c | c c c }

			 \textbf{Alias} & $w_0$ & $w_i$ & $q$ & $z_T$ &$H_0 [km/s Mpc^{-1}]$ & $\omega_c\equiv \Omega_ch^2$ & $\Omega_{m}^{(0)}$    
			 \\
			 \hline 
			(I)   $\Lambda CDM$-P  & -1 & -1 & 1 & 1 & 67.27& 0.1198 & 0.3156  \\
			(II)  $SEoS$-P &  -0.92 & -0.99 & 9.97 & 0.28 & 67.27&0.1198 &  0.3156  \\
			(III) $CPL$-P & -0.92 & -0.99 & 1 & 1 & 67.27& 0.1198 & 0.3156  \\
			(IV) $SEoS$-bf &  -0.92 & -0.99 & 9.97 & 0.28 & 73.22& 0.1568 &  0.3340  \\
		\end{tabular}
		\caption{Models used in this work. The cosmological parameters in models I-III correspond to Planck TT,TE,EE+lowP \citep{Ade:2015xua}, whereas model IV has the best fit values obtained in \citep{Jaber:2017bpx}. The rest of the parameters were kept fixed for all cases: $\Omega_bh^2 = 0.02225$, $\ln(10^{10}A_s) = 3.094$, $n_s = 0.9645$, also corresponding to those reported by Planck.
		} 
		\label{table:models}
	\end{center}
\end{table*}

\begin{figure}[t]
	\centering
	\includegraphics[width=\linewidth]{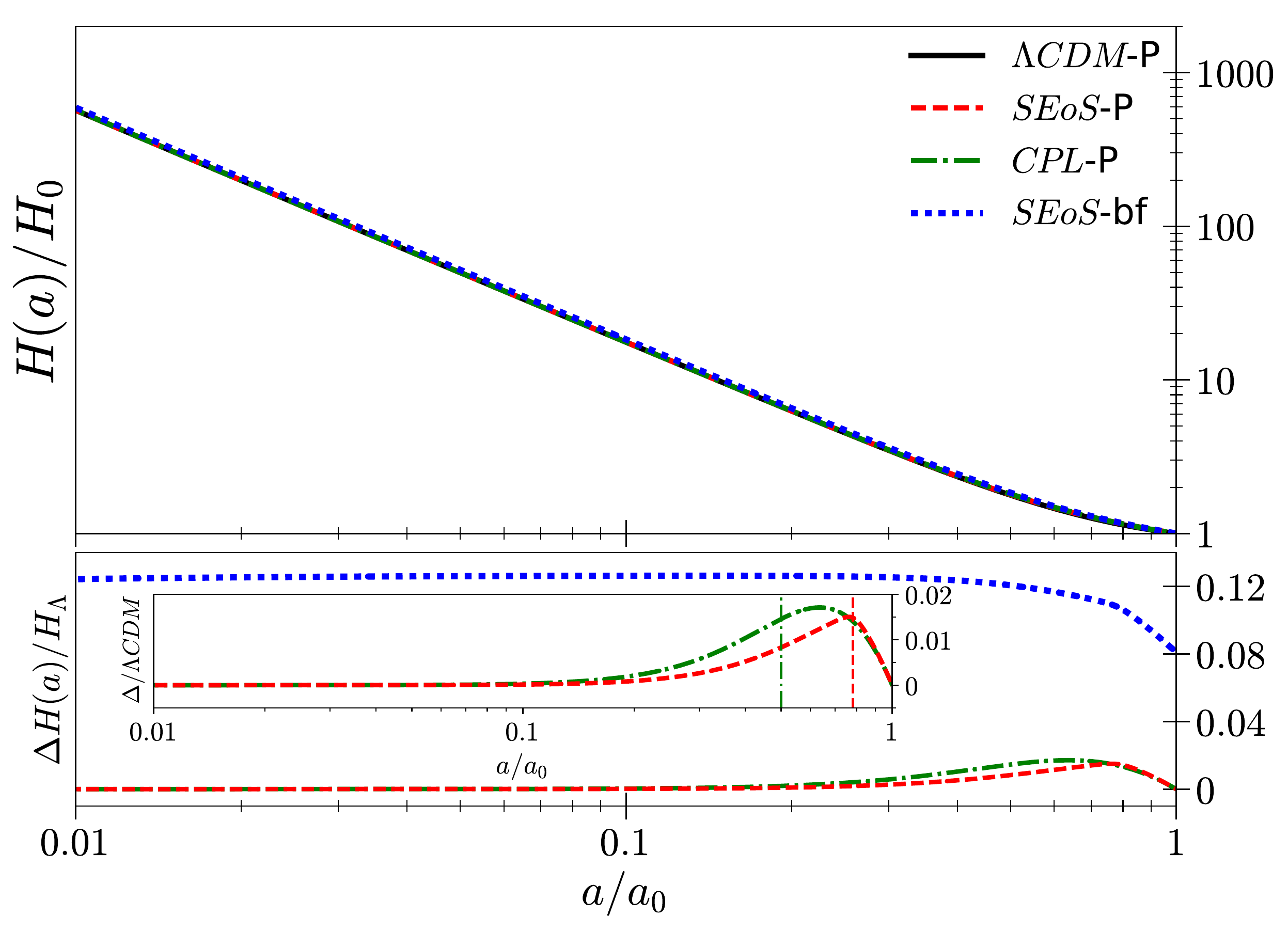}
	\caption{[Color online] (Upper panel) Hubble expansion rate normalized to $H_0$ for the models described in Table \ref{table:models}. (Bottom panel) Ratio of solutions for equation \eqref{eq:steepeos} to a cosmological constant: $\Delta H/H_{\La} \equiv (H-H_{\La})/H_{\La}$, where $H_{\La}$ refers to the solution $\La CDM$-P. The inset plot shows only the solutions $SEoS$-P and $CPL$-P compared to $\Lambda CDM$-P. The vertical lines represent the transition redshift for each model. }
	\label{fig:hubble} 
\end{figure}
For the free parameters in equation \eqref{eq:steepeos} we have chosen the best fit values obtained in \citep{Jaber:2017bpx} from the combination of BAO measurements \cite{Beutler:2011hx,Ross:2014qpa,Padmanabhan2pc,Alam:2016hwk,Delubac:2014aqe,Font-Ribera:2013wce} 
and the local determination of $H_0$ \citep{localhubble}. This corresponds to: $w_0=-0.92$, $w_i=-0.99$, $q=9.97$ and $z_T=0.28$. 
The cosmological parameters, $\Omega_m$ and $H_0$, were set equal to those  reported by the Planck collaboration \citep{Ade:2015xua}, so we can compare the discrepancy arising only from the different dynamics of DE (table \ref{table:models}, models I-III). This is  $\Omega_m = 0.3089$ and $H_0 = 67.74$ ($\omega_c=0.1198$). 

However, to take into account the full result obtained in \cite{Jaber:2017bpx}, we also set the values for  $\Omega_m$ and $H_0$ to those obtained with the constraining procedure reported previously. This corresponds to the model IV from table \ref{table:models} and values $\Omega_m = 0.3340$ and $H_0 = 73.22$ ($\omega_c = 0.1568$).  This value for $H_0$ corresponds to the one reported in \cite{localhubble}, which is known to be in  tension with the value extrapolated from Planck measurements of CMB. This will be have an impact in our analysis.

The rest of the cosmological parameters was fixed to the values from Planck TT,TE,EE+lowP \citep{Ade:2015xua} used in our previous analysis. In particular, we set the same primordial spectrum to focus on the effect of a late time dynamical dark energy component. This is, we set: $\ln(10^{10}A_s) = 3.094$, $n_s = 0.9645$, and  $\Omega_bh^2 = 0.02225$.

\subsection{\label{subsec:perturbs} Perturbative regime}

We examine the  growth of perturbations during the matter-DE domination era using ``SEoS'' (equation \eqref{eq:steepeos}) as the model for DE. 

For a late time Universe we have a mixture of matter and DE and we know radiation to be sub-dominant. 
In that case the growth of over-densities can be studied in the Newtonian limit of the formalism this is, considering non-relativistic components that are well inside the horizon. For coupled fluids we have:

\begin{widetext}
\begin{align}
	\label{eq:multi_m_de_a}
	a^2\frac{d^2\delta_i(a)}{da^2} + a\left(3+\frac{\dot H}{H^2}\right) \frac{d\delta_i(a)}{da} 
	- \left[\frac{3}{2} \Sigma_j (\Omega_j\delta_j) - \frac{(c_s^2)_i k^2}{a^2H^2}\delta_i(a) \right] = 0, \\
 	i,j = \text{matter}, \text{dark energy}. \nonumber 
\end{align}
\end{widetext}
where we have used $H^2 = \frac{8\pi G}{3}\bar\rho$. The density contrast of the i-th fluid is represented by 
$\delta_i\equiv (\rho_i-\bar{\rho})/\bar{\rho}$, 
where $\bar{\rho}$ is the background density, and $(c_s^2)_i$ represents the corresponding speed of sound, defined by $(c_s^2)_i\equiv\frac{\delta P_i}{\delta\rho_i}$.

We find the solutions for equation \eqref{eq:multi_m_de_a} in  the particular case of $\delta_{DE}=0$, this is, when DE does not cluster, since the spatial fluctuations of  typical dark energy models are very much suppressed with respect to those of dark matter.  

In addition to finding the numerical solutions of equation \eqref{eq:multi_m_de_a}, we also used a modified version of the Boltzmann solver \texttt{CAMB} \citep{Lewis:1999bs} in which we introduced ``SEoS'' as the background model.

\section{\label{sec:results}Results}

Regarding the solution for $w_{DE}(a)$ we choose to explore the different scenarios which are referenced in Table \ref{table:models} and were chosen as explained below: 

\begin{itemize}
    \item Model ``$\Lambda CDM$-P" refers to a cosmological constant scenario with $\Omega_m$ and $h$ fixed to Planck cosmology \citep{Ade:2015xua}. 
    \item Model ``$SEoS$-P''  refers to the best fit  values for the parameters in equation \eqref{eq:steepeos} as obtained in \citep{Jaber:2017bpx} while maintaining $\Omega_m$ and $h$ to a Planck cosmology \citep{Ade:2015xua}. 
    \item Model ``$CPL$-P'' refers to the scenario where we adopt the CPL limit of the above solution, meaning we keep   $\{w_0, w_i\} = \{-0.92, -0.99 \}$, as obtained in \citep{Jaber:2017bpx} and $\{\Omega_m, h\}$  fixed to a Planck cosmology \citep{Ade:2015xua}, but we make  $z_T$ = $q$ = 1. 
    \item Finally, Model ``$SEoS$-bf'' refers to the best fit values for the parameters in equation \eqref{eq:steepeos} (i.e. $\{w_0, w_i, q, z_T \} = \{-0.92, -0.99, 9.97, 0.28 \}$) with  $\Omega_m$ and $h$ also fixed to the best fit  values obtained in \citep{Jaber:2017bpx}.
\end{itemize}

The corresponding expansion histories for those models are shown in figure \ref{fig:hubble}, where we plot $H(a)/H_0$ and the relative ratio from models $SEoS$-P, $CPL$-P and $SEoS$-bf to $\La CMD$-P in the bottom panel: $\Delta H/H_{\La} \equiv (H-H_{\La})/H_{\La}$. 

\subsection{\label{subsec:growth} Growth function}
In the case where $\delta_{DE}=0$, equation \ref{eq:multi_m_de_a} reduces to:
\begin{equation}
    \label{eq:dm_a}
    a^2\frac{d^2\delta_m(a)}{da^2} + a\left(3+\frac{\dot H}{H^2}\right) \frac{d\delta_m(a)}{da} - \frac{3}{2}\Omega_m\delta_m(a)= 0.
\end{equation} 
This can be solved by setting initial conditions in the matter dominated era, $a_{ini} = 10^{-3}$, since we know that during this epoch, the solution for the growth function is $\delta_m(a) = a$, we have $\delta_m(a_{ini})=a_{ini} = 10^{-3}$ and $\delta_m'(a_{ini})=1$.

\begin{figure}[h]
	\centering
	\begin{subfigure}[b]{0.45\textwidth}
		\includegraphics[width=\textwidth]{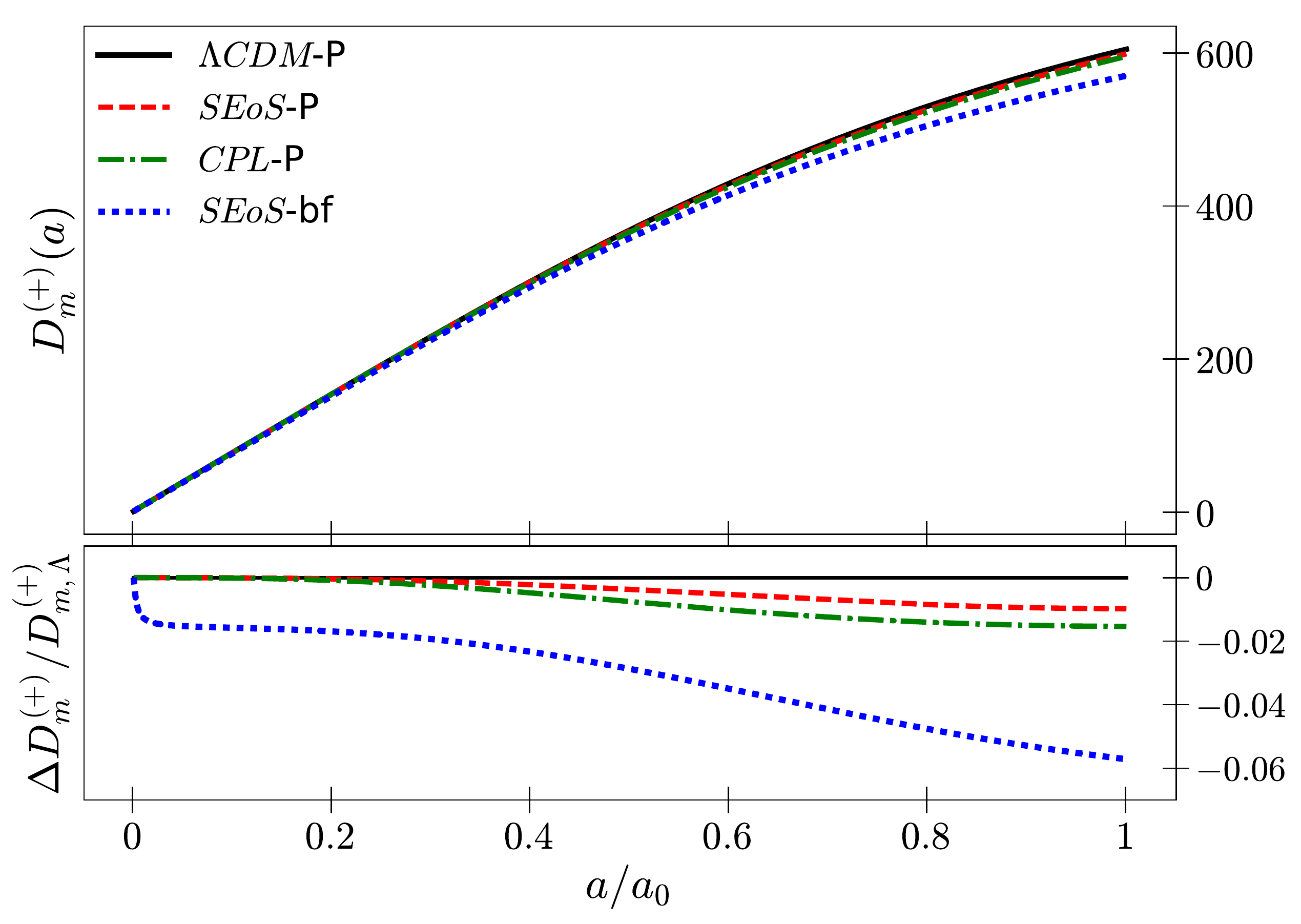}
		\caption{}
		\label{fig:Dm}
	\end{subfigure}

	\begin{subfigure}[b]{0.45\textwidth}
		\includegraphics[width=\textwidth]{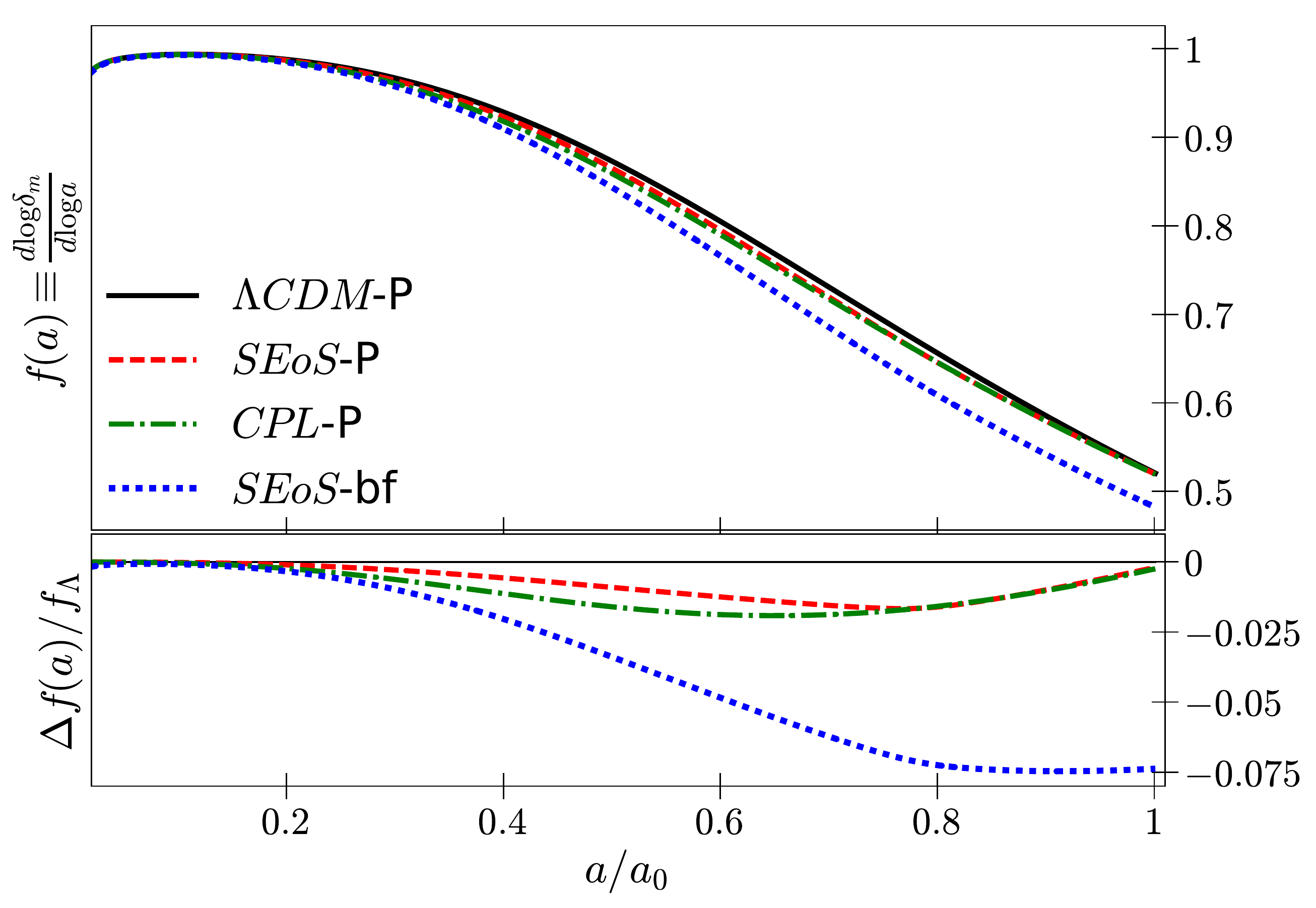}
		\caption{}
		\label{fig:fa}
	\end{subfigure}
	\caption{[Color online] \textbf{(a)}Evolution of matter overdensities normalized to the initial time, $D^{(+)}_m(a)$, and \textbf{(b)} logarithmic growth function,  $f(a)$ for the models in table \ref{table:models}. The bottom panel displays the relative difference to $\Lambda CDM$-P solution: $\Delta \mathcal{F} /\mathcal{F }_{\Lambda} \equiv \frac{\mathcal{F }- \mathcal{F }_{\Lambda}}{\mathcal{F }_{\Lambda}}$  with $\mathcal{F }$  representing $D^{(+)}_m(a)$ or $f(a)$, respectively. } 
	\label{fig:Dm-fa}
\end{figure}

A solution for equation \eqref{eq:dm_a} can be given up to a normalization. We choose to normalize it such that $D_m^{(+)}(a)=1$ at $a=a_{ini}$, so we enhance the differences arising at present time. This is shown in Figure \ref{fig:Dm} for the models under consideration. 
Once we have the solution to equation \eqref{eq:dm_a}, we can also find the logarithmic growth function, $f(a) \equiv \frac{d\log \delta_m(a)}{d\log a}$ (see  fig. \ref{fig:fa}).

To get a better idea of the effect of different dark energy models in the growth functions $D^{(+)}(a)\equiv\frac{\delta_m(a)}{\delta_m(a_{ini})}$, $f(a)$, we take the relative difference to a $\Lambda CDM-P$ scenario: $\Delta\mathcal{F}/\mathcal{F }_{\Lambda}$ $\equiv \frac{\mathcal{F}-\mathcal{F }_{\Lambda}}{\mathcal{F }_{\Lambda}}$  with $\mathcal{F}=\{D^{(+)}_m 
,f(a)\}$ and $\mathcal{F}_{\La}$ the solution assuming $\La CDM$-P as background model.
This is show, for instance, in the bottom panel of plots \ref{fig:Dm} and \ref{fig:fa}, respectively. 
%

Regarding the results for $D_m^{(+)}$ we find deviations from $\Lambda CDM$ that are of order:  
\begin{itemize}
   \item  of $1\%$ at $z=0$ if we assume model $SEoS$-P as our DE component,
   \item  of order $1.5\%$, for $CPL$-P,
   \item  and of order of $6\%$ at $z=0$  taking  $SEoS$-bf. 
\end{itemize}
The differences in $\Delta f(a)/f_{\La}$ are consistent, showing deviations at percent level: the fastest expansion rate corresponds to $SEoS$-bf model (as indicated in figure \ref{fig:hubble}), followed by $CPL$-P and $SEoS$-P. Hence, we obtained a slower growth of structure and a slower logarithmic growth rate in $SEoS$-bf model (followed by $CPL$-P and $SEoS$-P).

It is important to note that the discrepancy between $\Lambda CDM$ and a dynamic form of dark energy is bigger for the CPL scenario that the  case $\{w_0, w_i, q, z_T \} = \{-0.92, -0.99, 9.97, 0.28 \}$ (Model $CPL$-P versus Model $SEoS$-P in figure \ref{fig:Dm-fa}). This is due to the fact that the CPL limit  has $z_T = 1$, which implies that for  this case $w(z) \rightarrow -0.92$ for $z\leq1$, whereas Model $SEoS$-P has a later transition redshift, implying that $w(z) \rightarrow -0.92$ for $z\leq0.28$.

\subsection{\label{subsec:pk}Linear matter power spectrum}
By means of a modified version of \texttt{CAMB} \citep{Lewis:1999bs} in which we incorporated ``SEoS'' as expansion model and considered negligible DE perturbations, we computed the linear matter power spectrum, $P(k)$, which is calculated in the synchronous gauge, used internally by the code.

\subsubsection{SEoS: DE dynamics only}
\begin{figure}[h]
	\centering
	\begin{subfigure}[b]{0.45\textwidth}
		\includegraphics[width=\textwidth]{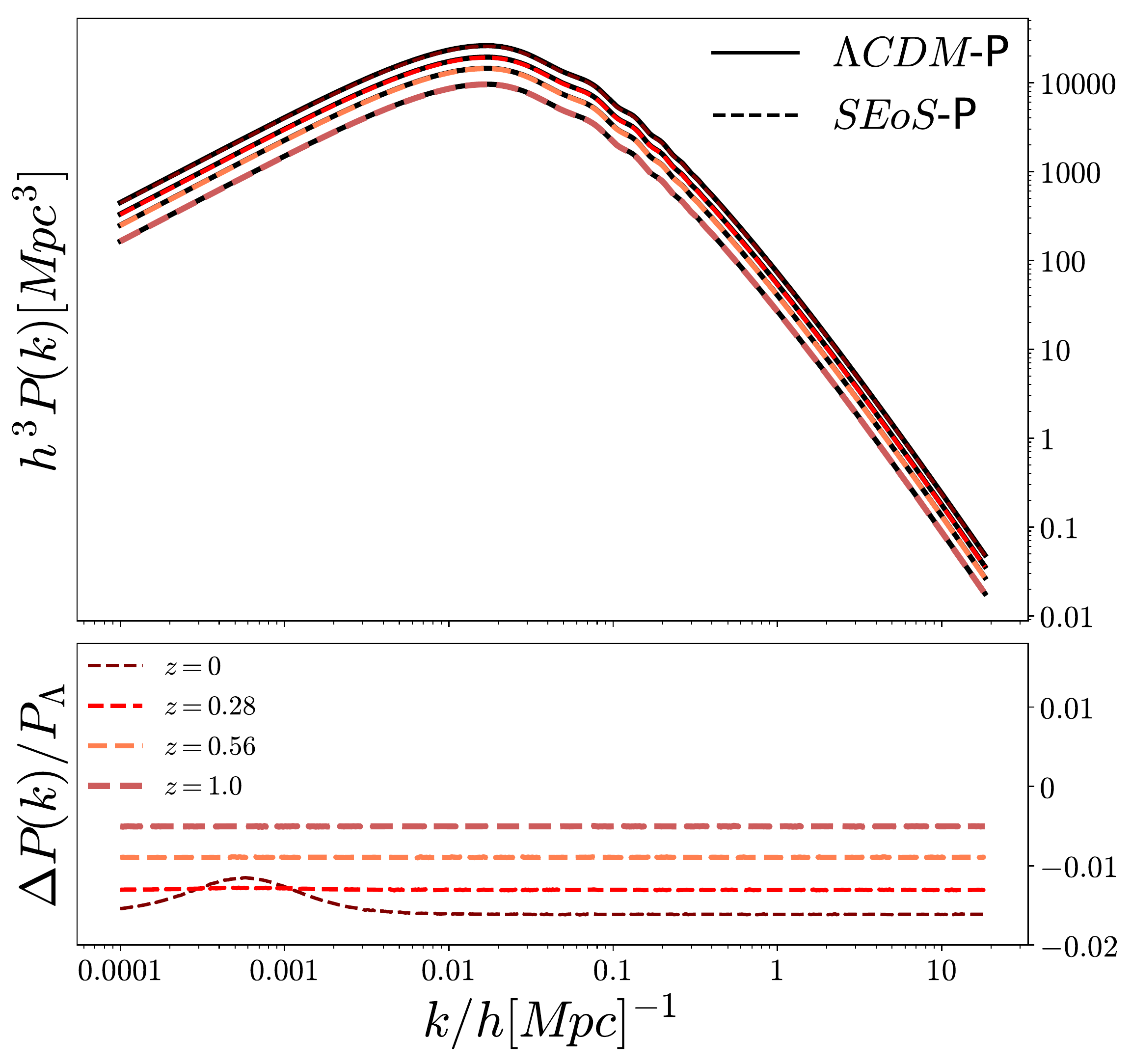}
		\caption{}
		\label{fig:pk}
	\end{subfigure}

	\begin{subfigure}[b]{0.45\textwidth}
		\includegraphics[width=\textwidth]{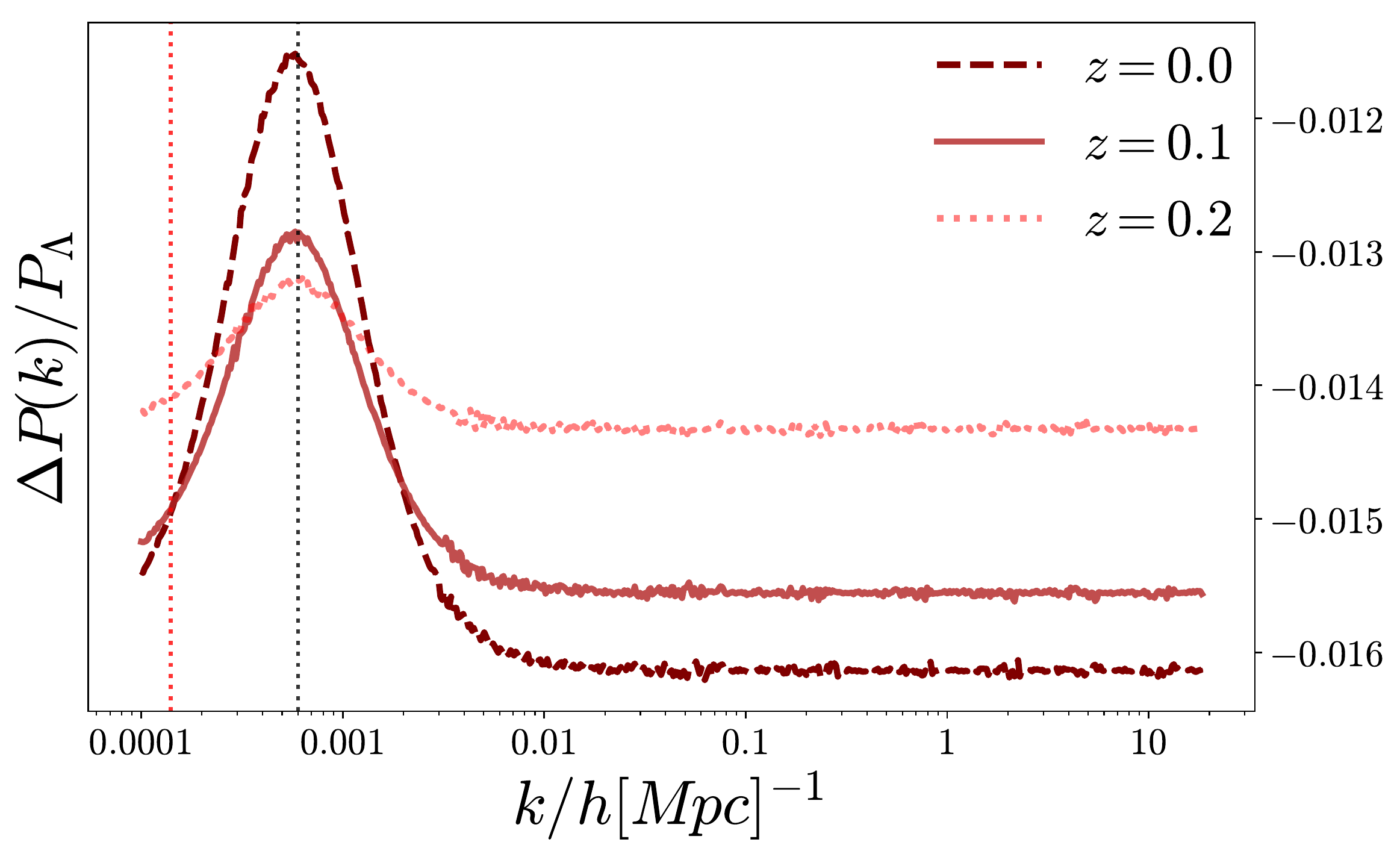}
		\caption{}
		\label{fig:pkzoom}
	\end{subfigure}
	\caption{[Color online] \textbf{(a)} Linear matter power spectra for Models I and  II  (Table \ref{table:models}) at different redshift values ($z=0$, $z=z_T=0.28$, $z=2z_T=0.56$, and $z=1$) and the ratio from ``$SEoS$-P'' to $\Lambda CDM$. 
	\textbf{(b)} Zoom-in of the previous plot showing $\Delta P(k)/P_{\Lambda}(k)$ for $z<z_T$: $z=0, 0.1, 0.2$.
	The (red) vertical line at $k=1.403\times10^{-4}h^{-1}/Mpc$ indicates the mode associated to the transition redshift ($z_T=0.28$), $k_T =a_TH(a_T)$, whereas the (black) vertical line in $k\approx6\times10^{-4}h^{-1}/Mpc$, indicates the maximum of the  bump in $\Delta P(k)/P_\La(k)$.} 
	
	\label{fig:pk1}
\end{figure}

Our results are shown in Figure \ref{fig:pk1}. In this we show the linear matter spectrum for $\Lambda CDM$ and $SEoS$-P (figure \ref{fig:pk}) and their ratio $\Delta P(k) /P_{\La} \equiv (P(k)- P_{\Lambda CDM}(k))/P_{\Lambda CDM}(k)$ for different redshift values ($z=0$, $z=z_T=0.28$, $z=2z_T=0.56$, and $z=1$). 

We notice a  decrease in amplitude for all Fourier modes, of order $0.5\%$ ($1.7\%$) for redshift values $z=1$ ($z=0$), respectively (see figure \ref{fig:pk}). This is to be expected since we have seen that a consequence of the dynamics of $SEoS$-P is an Universe that expands more rapidly as compared to one dominated by a cosmological constant. The effect  appears after the transition has occurred, since for $z>z_T$, our EoS behaves as a cosmological constant term ($w_i\approx-1$). 
In addition to this decrease, we notice a bump in $k\approx 6\times10^{-4}h^{-1}/Mpc$ for $\Delta P(k)/P_{\La}|_{z=0}$. This is better depicted in figure \ref{fig:pkzoom}, where we show the ratio between $SEoS$-P and $\La CDM$-P
for power spectra after the transition has occurred ($z<z_T$). From the bottom panel of  figure \ref{fig:pk} we notice the bump appears only after the transition has occurred, and in figure \ref{fig:pkzoom} we see it increases as $z \rightarrow 0$. 
We can know which modes are entering to the horizon during and after the transition epoch of $z_T=0.28$. \newline Using $h=0.6774$ we have $k_T\equiv a_T H(a_T) = 1.403\times10^{-4}h^{-1}/Mpc$ (shown as a red dotted vertical line in figure \ref{fig:pkzoom}) with $a_T=1/(1+z_T)$. Which means that modes $k<k_T$ enter into the horizon after the abrupt transition took place.

\begin{figure}[H]
	\centering
	\includegraphics[width=0.9\linewidth]{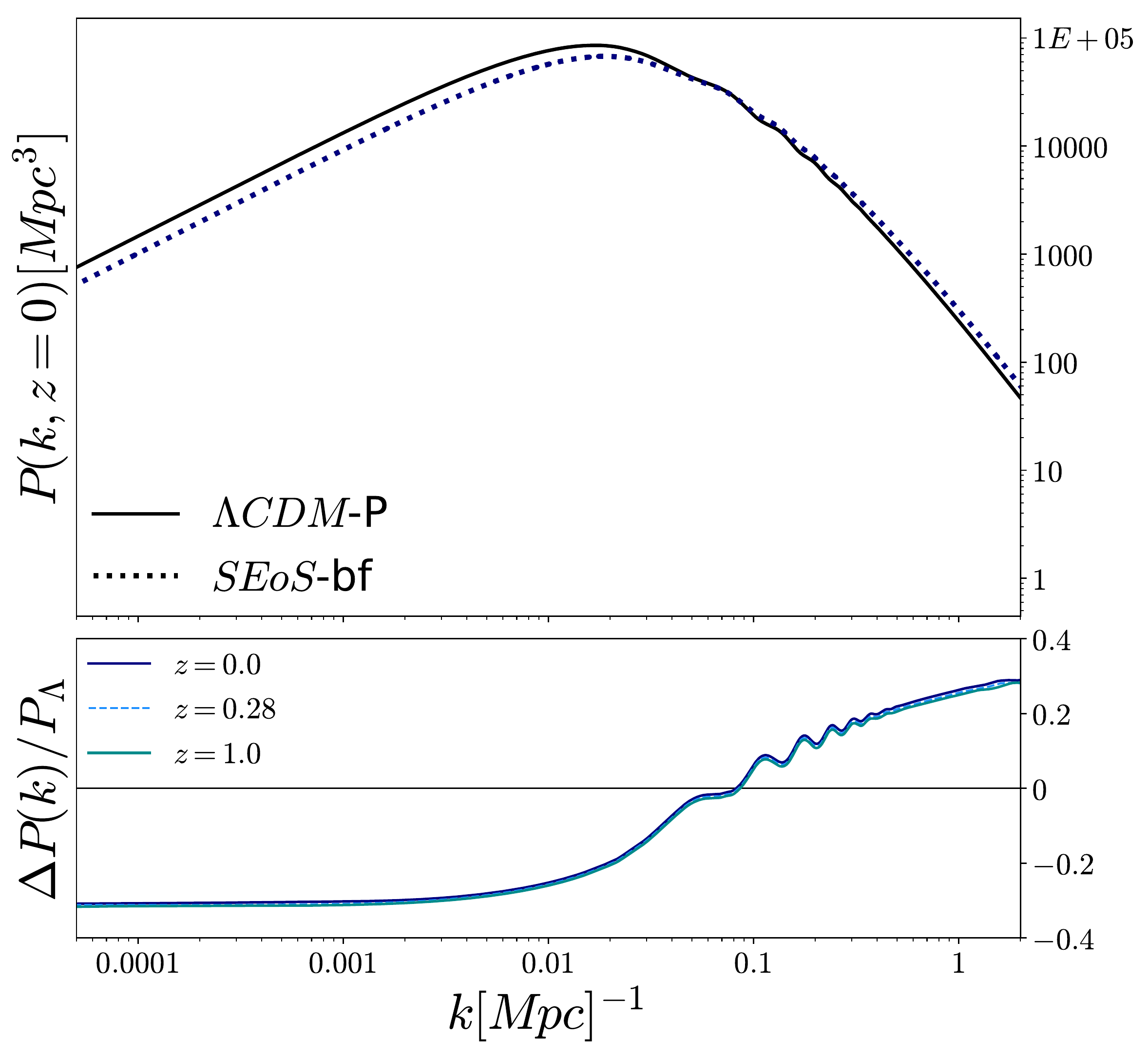}	
	\caption{[Color online] (Upper panel) Linear matter power spectra at present time for $SEoS$-bf and $\Lambda CDM$-P from Table \ref{table:models}. (Bottom panel) Ratio between $SEoS$-bf  to $\Lambda CDM$-P at different redshift values ($z=0$, $z=z_T=0.28$, $z=2z_T=0.56$, $z=1$).}
	\label{fig:pk_zss}.
\end{figure}

\subsubsection{SEoS: best fit value}

Now, in figure \ref{fig:pk_zss} we show the difference in power spectrum between $\Lambda CDM$ and the model $SEoS$-bf, in which not only the dynamics of DE is different but also the amount of matter, $\Omega_m^{(0)}$ and Hubble factor $H_0$. Notice that in this case we report $P(k)[Mpc^3]$ and $k[Mpc^{-1}]$ to take into account the fact that   each model has its corresponding $h$ value. It is important to recall that we have set the same initial power spectra for all our models, even in the $SEoS$-bf case.   

The bottom panel of figure \ref{fig:pk_zss} shows a decrease in power spectrum for small modes ($k\leq0.01/Mpc$), and a similar increase in amplitude for the biggest modes ($k\geq1/Mpc$).  Those modes ($k\geq1/Mpc$) entered first into the horizon and given that the Universe in $SEoS$-bf model expands more rapidly than in $\La CDM$-P (see figure \ref{fig:hubble}), they have had more time to evolve and accrete mass, hence, generating more power in the $SEoS$-bf power spectrum. 

It is customary to express the matter power spectrum at late times in terms of the initial power spectrum, the matter transfer function and the growth function \cite{Dodelson:book}:
\begin{equation}
	P(k, a) = 2\pi^2\delta_H^2\frac{k^n}{H_0^{n+3}}T^2(k)\left[\frac{D_1(a)}{D_1(a=1)}\right]^2,
\end{equation}
and since we know that the primordial power spectrum has been kept the same in all models tested, and the transfer function is roughly the same $(\approx9/10)$ for small modes, we can estimate the amount of deviation for small modes from $P_{SEoS}(k)/P_{\Lambda}$ to be of the order $\frac{(D_{1,seos}/D_{1,\Lambda})^2}{(H_{0,seos}/H_{0,\La})^4}$. From results in figures \ref{fig:hubble} and \ref{fig:Dm} we get $\frac{(D_{1,seos}/D_{1,\Lambda})^2}{(H_{0,seos}/H_{0,\La})^4} = \frac{(0.95)^2}{(1.08)^4} \approx 0.66 $ which in turns means $\Delta P(k) \approx -33\%$, in agreement with figure \ref{fig:pk_zss}.

\subsection{\label{subsec:lss} Large scale structure}

\begin{figure}[H]
	\centering
	\includegraphics[width=0.98\linewidth]{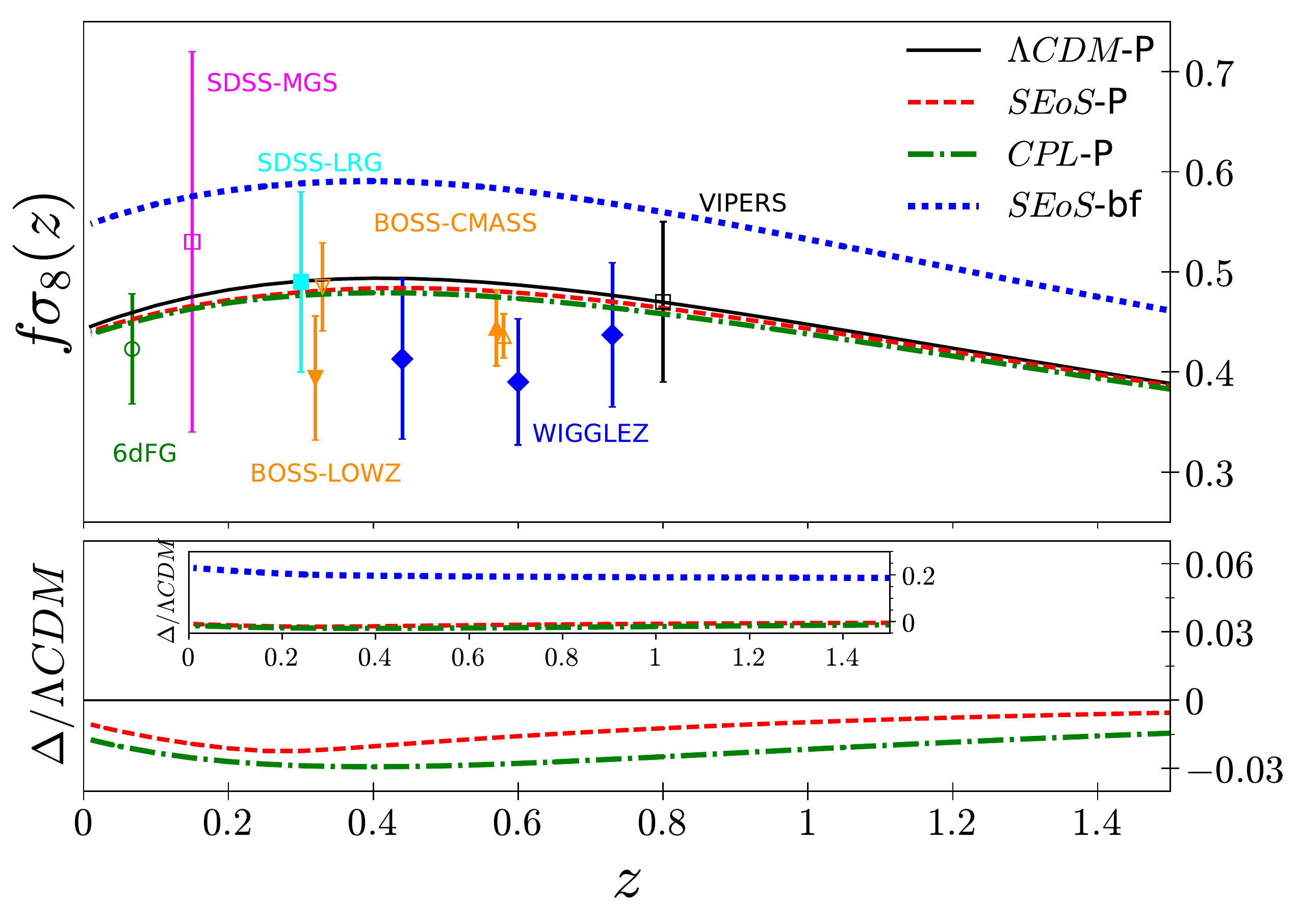}	
	\caption{[Color online] Predictions on $f\sigma_8(z)$ for   model of table \ref{table:models}. On top, we added the observational points reported by several surveys \cite{fs8.beutler,fs8.howlett,fs8:Oka:2013cba,Gil-Marin:2015sqa,fs8:Blake,fs8:delaTorre:2013rpa}. The bottom panel shows the relative difference with respect to $\La CDM$-P. }
	\label{fig:fsigma8}
\end{figure}

Galaxy redshift maps constrain the combination $f\sigma_8(z)$ using measurements of the redshift space distortions (RSD). So, in order to have an insight on the possible implications of the model into the growth of structure at large scales, we consider the $f\sigma_8(z)$ function and compare with some of the current observational constraints reported in the literature and listed below. This is shown in figure \ref{fig:fsigma8}.
We added the observational points reported by the following surveys: 6dFG \cite{fs8.beutler}, SDSS MGS \cite{fs8.howlett}, SDSS-LRG \cite{fs8:Oka:2013cba}, BOSS-LOWZ and BOSS-CMASS \cite{Gil-Marin:2015sqa}, WIGGLE-z  \cite{fs8:Blake} and the VIPERS \cite{fs8:delaTorre:2013rpa}.
As previously mentioned, we show the relative difference $\Delta\mathcal{F}/\mathcal{F }_{\Lambda}$ $\equiv \frac{\mathcal{F}-\mathcal{F }_{\Lambda}}{\mathcal{F }_{\Lambda}}$  with $\mathcal{F}$ = $\{f\sigma_8(z), dn/d\log M(z=0)\}$ and $\mathcal{F}_{\La}$ the solution assuming $\La CDM$-P as background model.

From this result we see that model $SEoS$-bf predicts a larger value for $f\sigma_8(z)$ at all redshift values $z\in[0,1.5]$. This increase (of order $\Delta f\sigma_8\sim 20\%$) makes model $SEoS$-bf in discordance with the current observations of $f\sigma_8(z)$, whereas $SEoS$-P and its $CPL$ limit are within observational error bars and deviate from $\La CDM$-P by less of $3\%$, in conformity with our previous results, in particular, we see that the difference in $\Delta f\sigma_8(z)$ is in agreement with the result shown in figure \ref{fig:fa}.

\begin{figure}[H]
	\centering
	\includegraphics[width=0.98\linewidth]{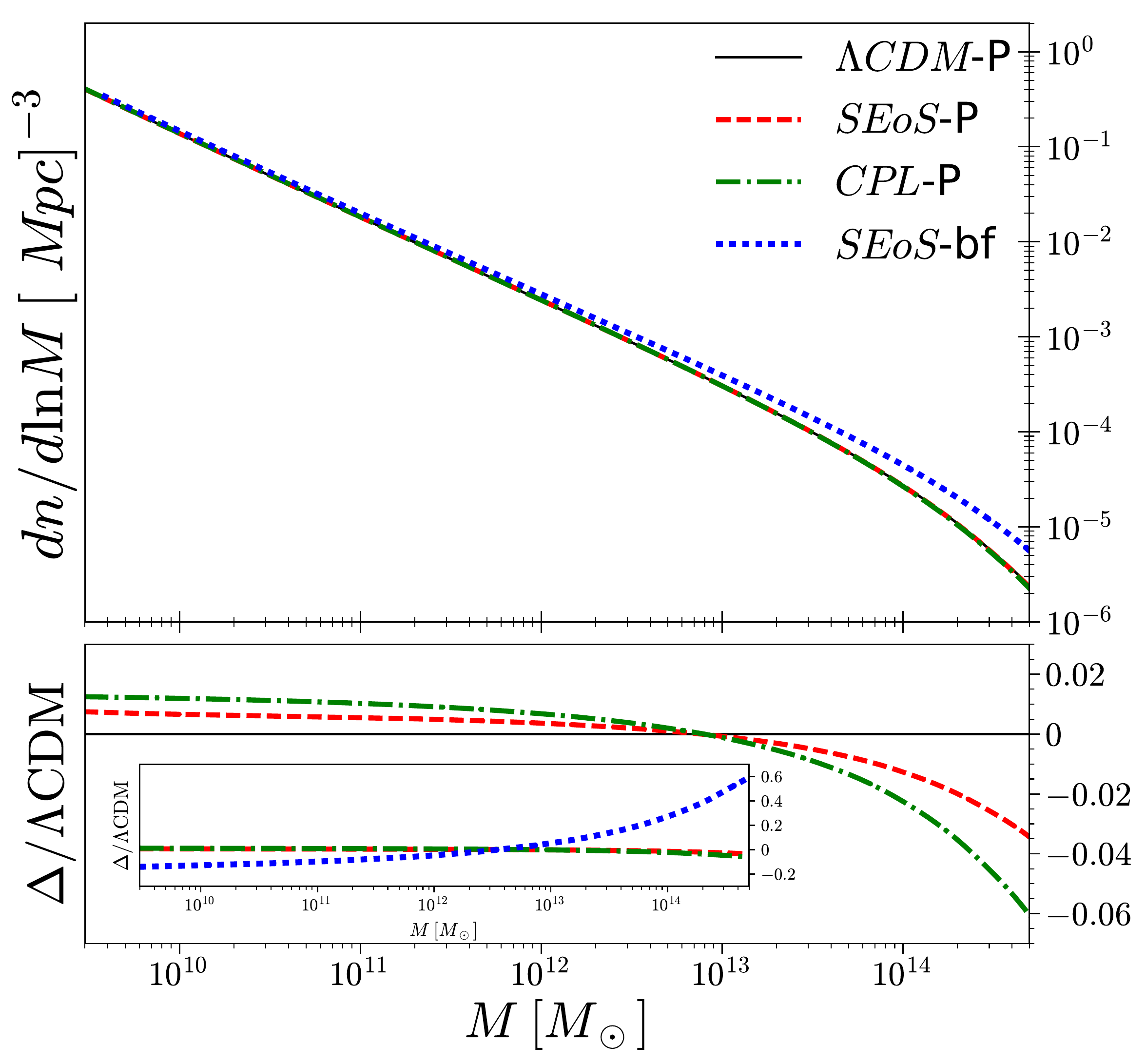}	
	\caption{[Color online] Differential mass function $dn/d\ln M$ for models of table \ref{table:models} at $z=0$. The bottom panel shows the relative difference to $\La CDM$-P.  }
	\label{fig:ps}.
\end{figure}

Additionally we consider the fractional number of collapsed structures by means of the Press-Schechter formalism, which describes the matter over-density field in real space by a smooth gaussian field whose variance on a sphere of radius R is $\sigma^2_R$ \citep{1974ApJ.PressSch}. In this formalism, the number of collapsed objects per unit volume with mass between $M$ and $M+dM$ is given by:

\begin{equation}
	\label{eq:ps}
	dn = -\sqrt{\frac{2}{\pi}}\frac{d\sigma_R}{dM}\left(\frac{\bar{\rho}_m\delta_{c}}{M\sigma_R^2}\right)\exp{\left(-\frac{\delta_c^2}{2\sigma^2_R}\right)}dM
\end{equation}
where $\delta_c=1.686$, the linear over-density at collapse is set to the value for $\La CDM$ since the dependence on cosmology is not strong \citep{pace}. In figure \ref{fig:ps} we show the differential mass function for the models considered and their relative ratio to $\La CDM$-P model. 

In this case we notice that $SEoS$-bf model predicts a decrease in the number of smalls structures (masses $M\sim\mathcal{O}(10^{10}M_{\odot})$) by $10\%$ compared to $\La CDM$ scenario, and an increase of $30\%$ ($M\sim\mathcal{O}(10^{14}M_{\odot})$) and as big as $50\%$ for the biggest collapsed structures ($M\geq4\times10^{14}M_{\odot}$). 

For $SEoS$-P and $CPL$-P models, the behavior is the opposite: we find an increase in the number of small objects (masses $M\leq6\times10^{12}M_{\odot}$) of order $1-2\%$ and a decrease in the number of big structures ($M\geq5\times10^{14}M_{\odot}$)  of $\sim 3\%$ for $SEoS$-P and $6\%$ for its $CPL$ limit.

We recall that the mass $M$ is inversely proportional to the wave-number since $M\propto R^3$ and $R=\pi/k$, indicating that large masses correspond to small modes (large scales) and vice versa. In $SEoS$-bf model, additionally to the DE dynamics we have a different value for $\bar{\rho}_m$ than in $\La CDM$ (see equation \eqref{eq:ps}), which impacts importantly the mass function, as we have just discussed. For models $SEoS$-P and $CPL$-P, however, the matter content is the same  and hence when we compare the differential mass function at a particular mass scale we are also comparing that function at the same mode. Moreover, since the age of the Universe is practically the same in models I-III (table \ref{table:models}), the resulting discrepancies previously discussed mean that large structures take more time to form in a $SEoS$ model, while small objects form more quickly. 

As a consequence we can say that we would expect to observe less massive galaxy clusters and more light structures (isolated galaxies and poorly populated clusters) in $SEoS$-P or $CPL$-P universe.

\section{\label{sec:conclusions} Conclusions}

We studied a DE model with the characteristic of a steep transition between two pivotal values. This model was previously analyzed at background level and its free parameters were tested against observations such as the latest local determination of $H_0$, the BAO peak and the angular distance to the CMB \citep{Jaber:2017bpx}, and constrained its free parameters to: ($w_0 = -0.92$, $w_i=-0.99$, $q=9.97$, $z_T=0.28$). 
This work investigates how a steep transition in the DE EoS can affect the growth of structure, and we restricted ourselves to the case of a smooth DE component. 

We find that the effect of a SEoS for DE in structure formation can basically be separated into two phenomena: \textbf{1)} On one hand the presence of a dynamical dark energy changes the expansion of the background, leading to different growth rates and affecting the matter fluctuations \textbf{2)} While on the other hand, the change in $\Omega_m$ as well the Hubble rate (according to the BFV obtained previously) has a bigger impact than just the DE dynamics, modifying the observable quantities such as $P(k)$, $f\sigma_8(z)$, and $dn/d\log M$ beyond the current observational constraints.

In the fist case we find that the change in the Hubble expansion \ref{fig:hubble} of $1.5\%$ percent at the transition epoch ($z_T=0.28$ or $z_T=1$ in ``$SEoS$-P'' or ``$CPL$-P'', respectively), impacts the growth functions in an equivalent amount, diminishing the growth of structure at linear order by $1.5\%-2\%$ (figure \ref{fig:Dm-fa}). This consequently imprints into $f\sigma_8(z)$ as a decrease of $\sim 2-3\%$ and lies in agreement with RSD observational measurements from surveys \cite{fs8.beutler,fs8.howlett,fs8:Oka:2013cba,Gil-Marin:2015sqa,fs8:Blake,fs8:delaTorre:2013rpa}. As for the differential mass function, $dn/d\log M(z=0)$, we find as a prediction, a slight increment in the number of small collapsed objects of order $1\%$ ($2\%$) and a decrement in the number of large structures or order $3\%$ ($6\%$) for $SEoS$-P ($CPL$-P) model. 

The $CPL$ limit of $SEoS$-P model (which means taking $z_T = q =1$ in equation \eqref{eq:steepeos}), consistently shows bigger differences from $\La CDM$ model than $SEoS$-P, as a result of an earlier (yet smooth) transition from $w_i \approx -1$ to a bigger value $w_0=-0.92$, which implies the DE dilutes first in $CPL$-P model. 
As for the matter power spectrum, we see that the change in expansion rate affects all Fourier modes equally, decreasing the amplitude of power spectrum  by $\sim 1.6\%$ at $z=0$. Additionally to this effect, we notice the appearance of a bump in the modes close to those entering near the steep transition, in the linear regime ($k\sim10^{-4}h^{-1}/Mpc$), which appears only after the transition took place ($z<z_T$) and increases amplitude as $z\longrightarrow0$. 
\newpage
In the second case, this is, for $SEoS$-P model, we find an interplay between having an Universe with  $\Delta\omega_c =  0.1568/0.1198 \sim 30 \% $ bigger than in a $\La CDM$ scenario, with the change in the expansion rate, such that the clustering is prevented at large scales (small Fourier modes or large masses) and enhanced at enhanced at small scales (large Fourier modes or small masses).  See for instance  figures \ref{fig:pk_zss} and \ref{fig:ps}. From the differential mass function, for instance, the prediction  is that the number of collapsed objects decreases (increases) by approximately $10\%$ ($50-60\%$) for light (the largest) structures. Lastly, the effect on $f\sigma_8(z)$, however implies that model $SEoS$-bf is not in agreement with RSD observational constraints. 

To summarize, the study of dynamics of Dark Energy is a matter of profound implications for our understanding of the Universe and its physical laws. 
Studying the behavior of a model beyond background level is nowadays required given the important amount of data coming from redshift galaxy surveys and its potential to test discrepancies among a cosmological constant, fluids with negative pressure or modifications to the gravity sector.
In this paper we have contributed towards that direction showing that the evolution of matter over-densities is sensitive to the parameters in equation \eqref{eq:steepeos}, and a model with a steep transition such as the one explored in this paper can lead to interesting features in the growth of structure.

\section*{Acknowledgements}

This project was done with funding from the CONACYT grant Fronteras de la Ciencia 000281 and PASPA-DGAPA UNAM. M.J. thanks Omar A. Rodr\'iguez L. for computational help and the group of Extragalactic Astronomy and Cosmology of Institute of Astronomy UNAM for fruitful discussions.

\bibliographystyle{unsrtnat}
\bibliography{SEoS_Perturbations}

\begin{thebibliography}{46}
\providecommand{\natexlab}[1]{#1}
\providecommand{\url}[1]{\texttt{#1}}
\expandafter\ifx\csname urlstyle\endcsname\relax
  \providecommand{\doi}[1]{doi: #1}\else
  \providecommand{\doi}{doi: \begingroup \urlstyle{rm}\Url}\fi

\bibitem[Weinberg(1989)]{RevModPhys.61.1}
Steven Weinberg.
\newblock The cosmological constant problem.
\newblock \emph{Rev. Mod. Phys.}, 61:\penalty0 1--23, Jan 1989.
\newblock \doi{10.1103/RevModPhys.61.1}.

\bibitem[Aghanim et~al.(2018)]{planck2018}
N.~Aghanim et~al.
\newblock {Planck 2018 results. VI. Cosmological parameters}.
\newblock 2018.

\bibitem[Scolnic et~al.(2018)]{Scolnic:2017caz}
D.~M. Scolnic et~al.
\newblock {The Complete Light-curve Sample of Spectroscopically Confirmed SNe
  Ia from Pan-STARRS1 and Cosmological Constraints from the Combined Pantheon
  Sample}.
\newblock \emph{Astrophys. J.}, 859\penalty0 (2):\penalty0 101, 2018.
\newblock \doi{10.3847/1538-4357/aab9bb}.

\bibitem[{Beutler} et~al.(2011){Beutler}, {Blake}, {Colless}, {Jones},
  {Staveley-Smith}, {Campbell}, {Parker}, {Saunders}, and
  {Watson}]{Beutler:2011hx}
F.~{Beutler}, C.~{Blake}, M.~{Colless}, D.~H. {Jones}, L.~{Staveley-Smith},
  L.~{Campbell}, Q.~{Parker}, W.~{Saunders}, and F.~{Watson}.
\newblock {The 6dF Galaxy Survey: baryon acoustic oscillations and the local
  Hubble constant}.
\newblock \emph{\mnras}, 416:\penalty0 3017--3032, October 2011.
\newblock \doi{10.1111/j.1365-2966.2011.19250.x}.

\bibitem[Ross et~al.(2015)Ross, Samushia, Howlett, Percival, Burden, and
  Manera]{Ross:2014qpa}
Ashley~J. Ross, Lado Samushia, Cullan Howlett, Will~J. Percival, Angela Burden,
  and Marc Manera.
\newblock {The clustering of the SDSS DR7 main Galaxy sample – I. A 4 per
  cent distance measure at $z = 0.15$}.
\newblock \emph{Mon. Not. Roy. Astron. Soc.}, 449\penalty0 (1):\penalty0
  835--847, 2015.
\newblock \doi{10.1093/mnras/stv154}.

\bibitem[Padmanabhan et~al.(2012)Padmanabhan, Xu, Eisenstein, Scalzo, Cuesta,
  Mehta, and Kazin]{Padmanabhan2pc}
Nikhil Padmanabhan, Xiaoying Xu, Daniel~J. Eisenstein, Richard Scalzo,
  Antonio~J. Cuesta, Kushal~T. Mehta, and Eyal Kazin.
\newblock A 2 per cent distance to z = 0.35 by reconstructing baryon acoustic
  oscillations – i. methods and application to the sloan digital sky survey.
\newblock \emph{Monthly Notices of the Royal Astronomical Society},
  427\penalty0 (3):\penalty0 2132--2145, 2012.
\newblock ISSN 1365-2966.
\newblock \doi{10.1111/j.1365-2966.2012.21888.x}.

\bibitem[Alam et~al.(2016)]{Alam:2016hwk}
Shadab Alam et~al.
\newblock {The clustering of galaxies in the completed SDSS-III Baryon
  Oscillation Spectroscopic Survey: cosmological analysis of the DR12 galaxy
  sample}.
\newblock \emph{Submitted to: Mon. Not. Roy. Astron. Soc.}, 2016.

\bibitem[Kazin et~al.(2014)]{Kazin:2014qga}
Eyal~A. Kazin et~al.
\newblock {The WiggleZ Dark Energy Survey: improved distance measurements to z
  = 1 with reconstruction of the baryonic acoustic feature}.
\newblock \emph{Mon. Not. Roy. Astron. Soc.}, 441\penalty0 (4):\penalty0
  3524--3542, 2014.
\newblock \doi{10.1093/mnras/stu778}.

\bibitem[Font-Ribera et~al.(2014)]{Font-Ribera:2013wce}
Andreu Font-Ribera et~al.
\newblock {Quasar-Lyman $\alpha$ Forest Cross-Correlation from BOSS DR11 :
  Baryon Acoustic Oscillations}.
\newblock \emph{JCAP}, 1405:\penalty0 027, 2014.
\newblock \doi{10.1088/1475-7516/2014/05/027}.

\bibitem[Delubac et~al.(2015)]{Delubac:2014aqe}
Timothée Delubac et~al.
\newblock {Baryon acoustic oscillations in the Lyman $\alpha$ Forest of BOSS
  DR11 quasars}.
\newblock \emph{Astron. Astrophys.}, 574:\penalty0 A59, 2015.
\newblock \doi{10.1051/0004-6361/201423969}.

\bibitem[Liang et~al.(2016)Liang, Zhao, Chuang, Kitaura, and
  Tao]{Liang:2015oqc}
Yu~Liang, Cheng Zhao, Chia-Hsun Chuang, Francisco-Shu Kitaura, and Charling
  Tao.
\newblock {Measuring Baryon Acoustic Oscillations from the clustering of
  voids}.
\newblock \emph{Mon. Not. Roy. Astron. Soc.}, 459\penalty0 (4):\penalty0
  4020--4028, 2016.
\newblock \doi{10.1093/mnras/stw884}.

\bibitem[Chevallier and Polarski(2001)]{Chevallier:2000qy}
Michel Chevallier and David Polarski.
\newblock {Accelerating universes with scaling dark matter}.
\newblock \emph{Int. J. Mod. Phys.}, D10:\penalty0 213--224, 2001.
\newblock \doi{10.1142/S0218271801000822}.

\bibitem[Linder(2003)]{Linder:2002et}
Eric~V. Linder.
\newblock {Exploring the expansion history of the universe}.
\newblock \emph{Phys. Rev. Lett.}, 90:\penalty0 091301, 2003.
\newblock \doi{10.1103/PhysRevLett.90.091301}.

\bibitem[Doran and Robbers(2006)]{Doran:2006kp}
Michael Doran and Georg Robbers.
\newblock {Early dark energy cosmologies}.
\newblock \emph{JCAP}, 0606:\penalty0 026, 2006.
\newblock \doi{10.1088/1475-7516/2006/06/026}.

\bibitem[Krauss et~al.(2007)Krauss, Jones-Smith, and
  Huterer]{KraussJonesHuterer2007}
Lawrence~M Krauss, Katherine Jones-Smith, and Dragan Huterer.
\newblock Dark energy, a cosmological constant, and type ia supernovae.
\newblock \emph{New Journal of Physics}, 9\penalty0 (5):\penalty0 141, 2007.

\bibitem[Linder(2006)]{Linder:2006ud}
Eric~V. Linder.
\newblock {Dark Energy in the Dark Ages}.
\newblock \emph{Astropart. Phys.}, 26:\penalty0 16--21, 2006.
\newblock \doi{10.1016/j.astropartphys.2006.04.004}.

\bibitem[Rubin et~al.(2009)]{Rubin:2008wq}
D.~Rubin et~al.
\newblock {Looking Beyond Lambda with the Union Supernova Compilation}.
\newblock \emph{Astrophys. J.}, 695:\penalty0 391--403, 2009.
\newblock \doi{10.1088/0004-637X/695/1/391}.

\bibitem[{Sollerman} et~al.(2009){Sollerman}, {M{\"o}rtsell}, {Davis},
  {Blomqvist}, {Bassett}, {Becker}, {Cinabro}, {Filippenko}, {Foley},
  {Frieman}, {Garnavich}, {Lampeitl}, {Marriner}, {Miquel}, {Nichol},
  {Richmond}, {Sako}, {Schneider}, {Smith}, {Vanderplas}, and
  {Wheeler}]{2009ApJ:703:1374S}
J.~{Sollerman}, E.~{M{\"o}rtsell}, T.~M. {Davis}, M.~{Blomqvist}, B.~{Bassett},
  A.~C. {Becker}, D.~{Cinabro}, A.~V. {Filippenko}, R.~J. {Foley},
  J.~{Frieman}, P.~{Garnavich}, H.~{Lampeitl}, J.~{Marriner}, R.~{Miquel},
  R.~C. {Nichol}, M.~W. {Richmond}, M.~{Sako}, D.~P. {Schneider}, M.~{Smith},
  J.~T. {Vanderplas}, and J.~C. {Wheeler}.
\newblock {First-Year Sloan Digital Sky Survey-II (SDSS-II) Supernova Results:
  Constraints on Nonstandard Cosmological Models}.
\newblock \emph{\apj}, 703:\penalty0 1374--1385, October 2009.
\newblock \doi{10.1088/0004-637X/703/2/1374}.

\bibitem[{Mortonson} et~al.(2010){Mortonson}, {Hu}, and
  {Huterer}]{2010PhRvD..81f3007M}
M.~J. {Mortonson}, W.~{Hu}, and D.~{Huterer}.
\newblock {Testable dark energy predictions from current data}.
\newblock \emph{\prd}, 81\penalty0 (6):\penalty0 063007, March 2010.
\newblock \doi{10.1103/PhysRevD.81.063007}.

\bibitem[Hannestad and Mortsell(2004)]{Hannestad:2004cb}
Steen Hannestad and Edvard Mortsell.
\newblock {Cosmological constraints on the dark energy equation of state and
  its evolution}.
\newblock \emph{JCAP}, 0409:\penalty0 001, 2004.
\newblock \doi{10.1088/1475-7516/2004/09/001}.

\bibitem[Jassal et~al.(2005)Jassal, Bagla, and Padmanabhan]{Jassal:2004ej}
H.~K. Jassal, J.~S. Bagla, and T.~Padmanabhan.
\newblock {WMAP constraints on low redshift evolution of dark energy}.
\newblock \emph{Mon. Not. Roy. Astron. Soc.}, 356:\penalty0 L11--L16, 2005.

\bibitem[Ma and Zhang(2011)]{Ma:2011nc}
Jing-Zhe Ma and Xin Zhang.
\newblock {Probing the dynamics of dark energy with novel parametrizations}.
\newblock \emph{Phys. Lett.}, B699:\penalty0 233--238, 2011.
\newblock \doi{10.1016/j.physletb.2011.04.013}.

\bibitem[Huterer and Turner(2001)]{Huterer:2000mj}
Dragan Huterer and Michael~S. Turner.
\newblock {Probing the dark energy: Methods and strategies}.
\newblock \emph{Phys. Rev.}, D64:\penalty0 123527, 2001.
\newblock \doi{10.1103/PhysRevD.64.123527}.

\bibitem[Weller and Albrecht(2002)]{Weller:2001gf}
Jochen Weller and Andreas Albrecht.
\newblock {Future supernovae observations as a probe of dark energy}.
\newblock \emph{Phys. Rev.}, D65:\penalty0 103512, 2002.
\newblock \doi{10.1103/PhysRevD.65.103512}.

\bibitem[Huang et~al.(2011)Huang, Bond, and Kofman]{Huang:2010zra}
Zhiqi Huang, J.~Richard Bond, and Lev Kofman.
\newblock {Parameterizing and Measuring Dark Energy Trajectories from
  Late-Inflatons}.
\newblock \emph{Astrophys. J.}, 726:\penalty0 64, 2011.
\newblock \doi{10.1088/0004-637X/726/2/64}.

\bibitem[Barboza and Alcaniz(2008)]{Barboza:2008rh}
E.~M. Barboza, Jr. and J.~S. Alcaniz.
\newblock {A parametric model for dark energy}.
\newblock \emph{Phys. Lett.}, B666:\penalty0 415--419, 2008.
\newblock \doi{10.1016/j.physletb.2008.08.012}.

\bibitem[Jaime et~al.(2018)Jaime, Jaber, and Escamilla-Rivera]{Jaime:2018ftn}
Luisa~G. Jaime, Mariana Jaber, and Celia Escamilla-Rivera.
\newblock {New parametrized equation of state for dark energy surveys}.
\newblock \emph{Phys. Rev.}, D98\penalty0 (8):\penalty0 083530, 2018.
\newblock \doi{10.1103/PhysRevD.98.083530}.

\bibitem[Blanton et~al.(2017)]{eboss}
Michael~R. Blanton et~al.
\newblock Sloan digital sky survey iv: Mapping the milky way, nearby galaxies,
  and the distant universe.
\newblock \emph{The Astronomical Journal}, 154\penalty0 (1):\penalty0 28, 2017.

\bibitem[collaboration(2016)]{DESIref}
DESI collaboration.
\newblock Desi final design report part i: Science,targeting, and survey
  design, 2016.
\newblock [Online; accessed 30-September-2018].

\bibitem[{LSST Science Collaboration} et~al.(2009){LSST Science Collaboration},
  {Abell}, {Allison}, {Anderson}, {Andrew}, {Angel}, {Armus}, {Arnett},
  {Asztalos}, {Axelrod}, and et~al.]{2009arXiv0912.0201L}
{LSST Science Collaboration}, P.~A. {Abell}, J.~{Allison}, S.~F. {Anderson},
  J.~R. {Andrew}, J.~R.~P. {Angel}, L.~{Armus}, D.~{Arnett}, S.~J. {Asztalos},
  T.~S. {Axelrod}, and et~al.
\newblock {LSST Science Book, Version 2.0}.
\newblock \emph{ArXiv e-prints, arXiv:0912.0201}, December 2009.

\bibitem[{Laureijs} et~al.(2011){Laureijs}, {Amiaux}, {Arduini},
  {Augu{\`e}res}, {Brinchmann}, {Cole}, {Cropper}, {Dabin}, {Duvet}, {Ealet},
  and et~al.]{2011arXiv1110.3193L}
R.~{Laureijs}, J.~{Amiaux}, S.~{Arduini}, J.~. {Augu{\`e}res}, J.~{Brinchmann},
  R.~{Cole}, M.~{Cropper}, C.~{Dabin}, L.~{Duvet}, A.~{Ealet}, and et~al.
\newblock {Euclid Definition Study Report}.
\newblock \emph{ArXiv e-prints, arXiv:1110.3193}, October 2011.

\bibitem[Jaber and de~la Macorra(2018)]{Jaber:2017bpx}
Mariana Jaber and Axel de~la Macorra.
\newblock {Probing a Steep EoS for Dark Energy with latest observations}.
\newblock \emph{Astropart. Phys.}, 97:\penalty0 130--135, 2018.
\newblock \doi{10.1016/j.astropartphys.2017.11.007}.

\bibitem[Mukherjee et~al.(2008)Mukherjee, Kunz, Parkinson, and Wang]{mukherjee}
Pia Mukherjee, Martin Kunz, David Parkinson, and Yun Wang.
\newblock Planck priors for dark energy surveys.
\newblock \emph{Phys. Rev. D}, 78:\penalty0 083529, Oct 2008.
\newblock \doi{10.1103/PhysRevD.78.083529}.

\bibitem[Ade et~al.(2016{\natexlab{a}})]{planck15DE}
P.~A.~R. Ade et~al.
\newblock {Planck 2015 results. XIV. Dark energy and modified gravity}.
\newblock \emph{Astron. Astrophys.}, 594:\penalty0 A14, 2016{\natexlab{a}}.
\newblock \doi{10.1051/0004-6361/201525814}.

\bibitem[Riess et~al.(2016)]{localhubble}
Adam~G. Riess et~al.
\newblock {A 2.4\% Determination of the Local Value of the Hubble Constant}.
\newblock \emph{Astrophys. J.}, 826\penalty0 (1):\penalty0 56, 2016.
\newblock \doi{10.3847/0004-637X/826/1/56}.

\bibitem[Ade et~al.(2016{\natexlab{b}})]{Ade:2015xua}
P.~A.~R. Ade et~al.
\newblock {Planck 2015 results. XIII. Cosmological parameters}.
\newblock \emph{Astron. Astrophys.}, 594:\penalty0 A13, 2016{\natexlab{b}}.
\newblock \doi{10.1051/0004-6361/201525830}.

\bibitem[Lewis et~al.(2000)Lewis, Challinor, and Lasenby]{Lewis:1999bs}
Antony Lewis, Anthony Challinor, and Anthony Lasenby.
\newblock {Efficient computation of CMB anisotropies in closed FRW models}.
\newblock \emph{Astrophys. J.}, 538:\penalty0 473--476, 2000.
\newblock \doi{10.1086/309179}.

\bibitem[Dodelson(2003)]{Dodelson:book}
Scott Dodelson.
\newblock \emph{{Modern cosmology}}.
\newblock Academic Press, San Diego, CA, 2003.

\bibitem[{Beutler} et~al.(2012){Beutler}, {Blake}, {Colless}, {Jones},
  {Staveley-Smith}, {Poole}, {Campbell}, {Parker}, {Saunders}, and
  {Watson}]{fs8.beutler}
Florian {Beutler}, Chris {Blake}, Matthew {Colless}, D.~Heath {Jones}, Lister
  {Staveley-Smith}, Gregory~B. {Poole}, Lachlan {Campbell}, Quentin {Parker},
  Will {Saunders}, and Fred {Watson}.
\newblock {The 6dF Galaxy Survey: z{\ensuremath{\approx}} 0 measurements of the
  growth rate and {\ensuremath{\sigma}}$_{8}$}.
\newblock \emph{\mnras}, 423\penalty0 (4):\penalty0 3430--3444, Jul 2012.
\newblock \doi{10.1111/j.1365-2966.2012.21136.x}.

\bibitem[Howlett et~al.(2015)Howlett, Ross, Samushia, Percival, and
  Manera]{fs8.howlett}
Cullan Howlett, {Ashley J.} Ross, Lado Samushia, {Will J.} Percival, and Marc
  Manera.
\newblock The clustering of the sdss main galaxy sample ii: mock galaxy
  catalogues and a measurement of the growth of structure from redshift space
  distortions at z=0.15.
\newblock \emph{MNRAS}, 449\penalty0 (1):\penalty0 848--866, 5 2015.
\newblock ISSN 0035-8711.
\newblock \doi{10.1093/mnras/stu2693}.

\bibitem[Oka et~al.(2014)Oka, Saito, Nishimichi, Taruya, and
  Yamamoto]{fs8:Oka:2013cba}
Akira Oka, Shun Saito, Takahiro Nishimichi, Atsushi Taruya, and Kazuhiro
  Yamamoto.
\newblock {Simultaneous constraints on the growth of structure and cosmic
  expansion from the multipole power spectra of the SDSS DR7 LRG sample}.
\newblock \emph{Mon. Not. Roy. Astron. Soc.}, 439:\penalty0 2515--2530, 2014.
\newblock \doi{10.1093/mnras/stu111}.

\bibitem[Gil-Marín et~al.(2016)]{Gil-Marin:2015sqa}
Héctor Gil-Marín et~al.
\newblock {The clustering of galaxies in the SDSS-III Baryon Oscillation
  Spectroscopic Survey: RSD measurement from the LOS-dependent power spectrum
  of DR12 BOSS galaxies}.
\newblock \emph{Mon. Not. Roy. Astron. Soc.}, 460\penalty0 (4):\penalty0
  4188--4209, 2016.
\newblock \doi{10.1093/mnras/stw1096}.

\bibitem[{Blake} et~al.(2010){Blake}, {Brough}, {Colless}, {Couch}, {Croom},
  {Davis}, {Drinkwater}, {Forster}, {Glazebrook}, and {Jelliffe}]{fs8:Blake}
Chris {Blake}, Sarah {Brough}, Matthew {Colless}, Warrick {Couch}, Scott
  {Croom}, Tamara {Davis}, Michael~J. {Drinkwater}, Karl {Forster}, Karl
  {Glazebrook}, and Ben {Jelliffe}.
\newblock {The WiggleZ Dark Energy Survey: the selection function and z = 0.6
  galaxy power spectrum}.
\newblock \emph{\mnras}, 406\penalty0 (2):\penalty0 803--821, Aug 2010.
\newblock \doi{10.1111/j.1365-2966.2010.16747.x}.

\bibitem[de~la Torre et~al.(2013)]{fs8:delaTorre:2013rpa}
S.~de~la Torre et~al.
\newblock {The VIMOS Public Extragalactic Redshift Survey (VIPERS). Galaxy
  clustering and redshift-space distortions at z=0.8 in the first data
  release}.
\newblock \emph{Astron. Astrophys.}, 557:\penalty0 A54, 2013.
\newblock \doi{10.1051/0004-6361/201321463}.

\bibitem[{Press} and {Schechter}(1974)]{1974ApJ.PressSch}
W.~H. {Press} and P.~{Schechter}.
\newblock {Formation of Galaxies and Clusters of Galaxies by Self-Similar
  Gravitational Condensation}.
\newblock \emph{\apj}, 187:\penalty0 425--438, February 1974.
\newblock \doi{10.1086/152650}.

\bibitem[Pace et~al.(2010)Pace, Waizmann, and Bartelmann]{pace}
F.~Pace, J.-C. Waizmann, and M.~Bartelmann.
\newblock {Spherical collapse model in dark-energy cosmologies}.
\newblock \emph{Monthly Notices of the Royal Astronomical Society},
  406\penalty0 (3):\penalty0 1865--1874, 08 2010.
\newblock ISSN 0035-8711.
\newblock \doi{10.1111/j.1365-2966.2010.16841.x}.

\end{thebibliography}

\end{document}